\documentclass[a4paper,11pt]{article}
\usepackage[utf8]{inputenc}
\usepackage[T1]{fontenc}
\usepackage{graphicx}
\usepackage{grffile}
\usepackage{longtable}
\usepackage{wrapfig}
\usepackage{rotating}
\usepackage[normalem]{ulem}
\usepackage{amsmath}
\usepackage{textcomp}
\usepackage{amssymb}
\usepackage{capt-of}
\usepackage{hyperref}
\usepackage{listings}
\usepackage{tcolorbox}
\usepackage{subcaption}
\usepackage{libertine}
\usepackage{lastpage}
\usepackage{a4wide}
\usepackage{alltt}
 \PassOptionsToPackage{hyphens}{url}\usepackage{hyperref}
\newcommand{\swhurl}[1]{https://archive.softwareheritage.org/#1}
\newcommand{\swhref}[2]{\href{\swhurl{#1}}{#2}}
\newcommand{\swhidref}[1]{\swhref{#1}{\small #1}}

\definecolor{dkgreen}{rgb}{0,0.6,0}
\definecolor{gray}{rgb}{0.5,0.5,0.5}
\definecolor{mauve}{rgb}{0.58,0,0.82}
\author{Roberto Di Cosmo\\Inria, Software Heritage, University of Paris, France\\roberto@dicosmo.org}
\date{September 2019}
\title{How to use Software Heritage for archiving and referencing your source code: guidelines and walkthrough}
\hypersetup{
 pdfauthor={dicosmo},
 pdftitle={How to use Software Heritage for archiving and referencing your source code: guidelines and walkthrough},
 pdfkeywords={},
 pdfsubject={},
 pdfcreator={Emacs 26.1 (Org mode 9.1.14)}, 
 pdflang={English}}
\begin{document}

\maketitle
Software source code is \emph{an essential research output}, and there is a growing
general awareness of its importance for supporting the research process
\cite{Borgman2012,Stodden-reprod-2012,Hinsen2013}.  Many research communities
strongly encourage making the source code of the artefact available by archiving
it in publicly-accessible long-term archives. Some have even put in place
mechanisms to assess research software, like the \emph{Artefact Evaluation} process introduced
in 2011 and now widely adopted by many computer science conferences \cite{Dagstuhl-Artefacts-2016},
and the \emph{Artifact Review and Badging} program of the ACM \cite{AcmBadges}.\\

Software Heritage \cite{swhipres2017,swhcacm2018} is a non profit, long term
universal archive specifically designed for software source code, and able to
store not only a software artifact, but \emph{also its full development history}. 
It provides the ideal place to \emph{preserve research software
artifacts}, and offers powerful mechanisms to \emph{enhance research articles} with
precise references to relevant fragments of your source code.\\

Using Software Heritage for your research software artifacts is straightforward
and involves three simple steps, described in the picture below:

\begin{center}
\begin{center}
\includegraphics[width=.8\textwidth]{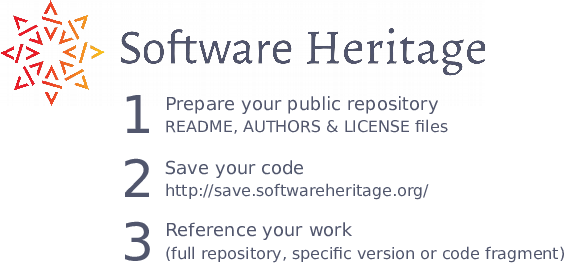}
\end{center}
\end{center}

In this document we will go through each of these three steps, providing
guidelines for making the most out of Software Heritage for your research:
Section \ref{sec:prepare} describes the best practices for preparing your source
code for archival; Section \ref{sec:save} shows how to archive your code in
Software Heritage; Section \ref{sec:walkthrough} shows the rich functionalities
you can use for referencing in your article source code archived in Software
Heritage; finally, in the Appendix you will find a formal description of the
different kinds of identifiers available for adressing the content archived in
Software Heritage.
\section{Prepare your repository}
\label{sec:orgdaa7897}
\label{sec:prepare}

We assume that your source code is hosted on a repository publicly accessible
(Github, Bitbucket, a GitLab instance, an institutional software forge, etc.)
using one of the version control systems supported by Software Heritage,
currently Subversion, Mercurial and Git
\footnote{For up to date information, see \url{https://archive.softwareheritage.org/browse/origin/save/}}.\\

It is highly recommended that you provide, in your source code repository,
appropriate information on your research artifact: it will make it more
appealing and useful to future users (which might actually be \emph{you} in a few
months).

Well established best practice is to include, at the toplevel of your
source code tree, three \emph{key files}, README, AUTHORS and LICENSE, with
the information described below.

\begin{description}
\item[{README}] : A description of the software.\\
This file should contain \emph{at least}
\begin{itemize}
\item the name of the software/project
\item a brief description of the project.
\end{itemize}
It is also \emph{highly recommended} to add the following information
\begin{itemize}
\item pointers to the project website and documentation,
\item pointer to the project development platform,
\item license for the project (if not in a separate LICENSE file),
\item contact and support information,
\item build/installation instructions or a pointer to a file containing them (usually INSTALL)
\end{itemize}
In could be useful to provide here also some information for the users, like a list of features or informations on how to use the source code
\end{description}

\begin{description}
\item[{AUTHORS}] : The list of all authors that need to be credited for the current version.\\
If you want to specify the role of each contributor in this list, we suggest
to use the taxonomy of contributors presented in \cite{gtinria2009}, which distinguishes
the following roles: \emph{Design, Architecture, Coding, Testing, Debugging, Documentation, Maintenance, Support, Management}.
\item[{LICENSE}] : The project license terms.\\
For Open Source Licenses, it is strongly recommended to use the standard names that can be found on the \url{https://spdx.org/licenses/} website.
\end{description}

Future users that find your artifact useful might want to give you credit by
citing it.  To this end, you might want to provide instructions on how you
prefer your artifact to be cited. There are many possibilities for doing
that, and you might want to also provide structured citation information
in specific formats, like CodeMeta (usually in a file named \textbf{Codemeta.json2})
or CFF (usually in a file named \textbf{CITATION.cff3}).

\subsection{Learning more}
\label{sec:orgf518040}
The seminal article \emph{Software Release Practice HOWTO} by E. S. Raymond
\cite{raymond2013} documents best practices and conventions for releasing
software that have been well established for decades, and form the basis
of most current recommendations. Interesting more recent resources include
the REUSE website \cite{reuse}, which provides detailed guidance
and tools to verify compliance with the guidelines, as well as \cite{SSI2018}, which focuses more on research software.

\section{Save your code}
\label{sec:org4782bbf}
\label{sec:save}

\begin{figure}[h!]
\begin{center}
\includegraphics[width=\textwidth]{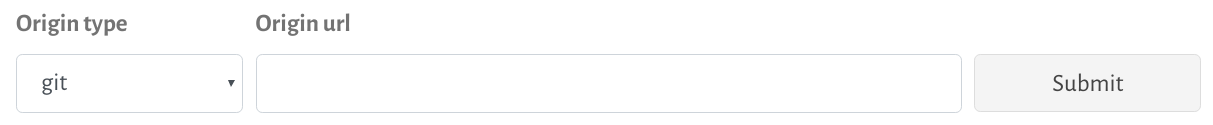}
\end{center}
\caption{The Save Code Now form}\label{fig:savecodenow}
\end{figure}

\noindent Once your code repository has been properly prepared, you only need to:

\begin{itemize}
\item go to \url{https://archive.softwareheritage.org/browse/origin/save/},
\item pick your version control system in the drop-down list, enter the code repository url
\footnote{Make sure to use the clone/checkout url as given by the development platform hosting your code. It can easily be found in the web interface of the development platform.},
\item click on the Submit button (see Figure \ref{fig:savecodenow}).
\end{itemize}

\noindent \textbf{That's it, it's all done!} No need to create an account or to provide
personal information of any kind. If the url you provided is correct, Software
Heritage will archive your repository, with its full development history,
shortly after. If your repository is hosted on one of the major forges we
already know, this process will take just a few hours; if you point to a
location we never saw before, it can take longer, as we will need to manually
approve it.\\

\noindent \textbf{For hackers:} you can also request archival programmatically, using the
Software Heritage API
\footnote{For details, see \url{https://archive.softwareheritage.org/api/1/origin/save/}};
this can be quite handy to integrate, for example, into a Makefile.

\section{Reference your work}
\label{sec:org1ac9e49}
\label{sec:walkthrough}

Once your source code has been archived, there are many ways to reference it in
your article. We present here three common use cases:
\begin{itemize}
\item link to the \emph{full repository} archived in Software Heritage,
\item link to a \emph{precise version of the software project},
\item link to a \emph{precise version of a source code file}, down to the level of the line of code.
\end{itemize}

To make this concrete, in what follows we use as a running example the article
\emph{A ``minimal disruption'' skeleton experiment: seamless map and reduce embedding
in OCaml} by Marco Danelutto and Roberto Di Cosmo \cite{Parmap2012} published
in 2012. This article introduced a nifty library for multicore parallel
programming that was distributed via the \url{gitorious.org} collaborative
development platform, at \url{gitorious.org/parmap}. Since Gitorious has been
shut down a few years ago, like Google Code and CodePlex, this example is
particularly fit to show why pointing to an \emph{archive} that has your code is
better than pointing to the collaborative development platform where you
developed it.

\subsection{Full repository}
\label{sec:org3fb889d}
In Software Heritage, we keep track of all the \emph{origins} from which source
code has been retrieved, and finding a given \texttt{origin} is as easy as
adding in front of it the prefix
\url{https://archive.softwareheritage.org/browse/origin}

These origins are the exact \emph{URLs of the version control system} that a
developer would use to clone a working repository, and are the same urls that
you pass to the \emph{Save Code Now} form described in Section \ref{sec:save}.

In our running example, for the Parmap code on \emph{gitorious.org}, this origin
is \url{https://gitorious.org/parmap/parmap.git}, so the URL of the \emph{persistently
archived full repository} is the following:

\url{https://archive.softwareheritage.org/browse/origin/https://gitorious.org/parmap/parmap.git}

Just add this link to your article, and your readers will be able to get hold of
the archived copy of your repository even if/when the original development
platform goes away (as it has actually happened for \texttt{gitorious.org} that
has been shut down in 2015).

Your readers can then browse the contents of your repository extensively,
delving into its development history, and/or directory structure, down to each
single source code file
\footnote{For a guided tour see \url{https://www.softwareheritage.org/2018/09/22/browsing-the-software-heritage-archive-a-guided-tour/}}.

\textbf{N.B.}: if you are unsure about what is the actual origin URL of your
repository, you can look it up using the search box that is available at
\url{https://archive.softwareheritage.org/browse/search/}

\subsection{Specific version}
\label{sec:org56af571}

Pointing to the full archived repository is nice, but a version controlled
repository usually contains all the history of development of the source code,
whiche records different states of the project, usually called \emph{revisions}.

In order to support reproducibility of scientific results, we need to be able to
pinpoint precisely the state(s) of the source code used in the article.
Software Heritage provides a very easy means of pointing to a precise
\emph{revision}, via a standard identifier schema, called SWH-ID, which is
\href{https://docs.softwareheritage.org/devel/swh-model/persistent-identifiers.html}{fully documented online} and is discussed in the article \cite{swhipres2018}.

In our running example, the Parmap article, the exact revision of the source
code of the library used therein has the following SWH-ID:

\begin{tcolorbox}
swh:1:rev:0064fbd0ad69de205ea6ec6999f3d3895e9442c2;\\
origin=https://gitorious.org/parmap/parmap.git;
\end{tcolorbox}

And you can turn this identifier into a clickable URL by prepending to it
the prefix \url{https://archive.softwareheritage.org/}: you can try it
live right now by clicking on \href{https://archive.softwareheritage.org/swh:1:rev:0064fbd0ad69de205ea6ec6999f3d3895e9442c2;origin=https://gitorious.org/parmap/parmap.git}{this link}.

\subsubsection{Getting your SWH-ID}
\label{sec:org4092446}
\label{ssec:getswhid}
\begin{figure}[h]
  \centering
  \includegraphics[width=\linewidth]{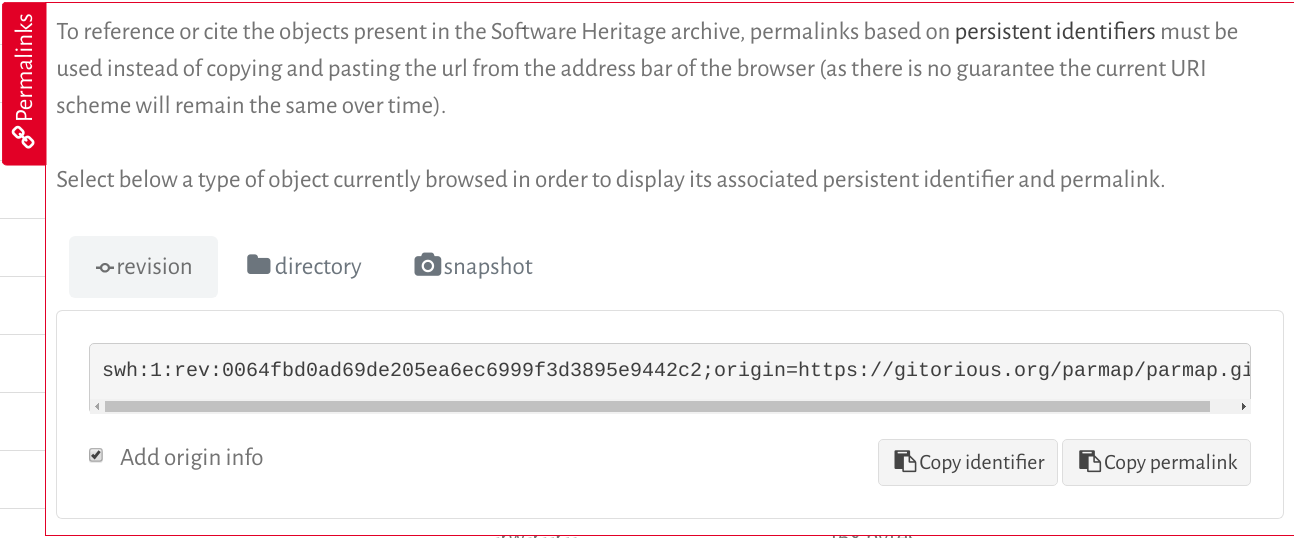}
  \caption{Obtaining a Software Heritage identifier using the permalink box on
    the archive Web user interface}
  \label{fig:permalink}
\end{figure}

 A very simple way of getting the right SWH-ID is to browse your archived code in
 Software Heritage, and to navigate to the revision you are interested in. Click
 then on the \emph{permalinks vertical red tab} that is present on all pages of the
 archive, and in the tab that opens up you select the \emph{revision} identifier:
 an example is shown in Figure \ref{fig:permalink}.

The two convenient buttons on the botton right allow you to copy the identifiers or the full permalink
in the clipboard, to insert in your article as you see fit.

\subsubsection{Generating and verifying SWH-IDs (for the geeks)}
\label{sec:org6680ada}
Version 1 of the SWH-IDs uses git-compatible hashes, so if you are using git as
a version control system, you can create the right SWH-ID by just prepending
\texttt{swh:1:rev:} to your commit hash. This might come pretty handy if you
plan to automate the generation of the identifiers to be included in your
article: you will always have your code and your article in sync!

Software Heritage identifiers can also be generated and verified independently
by anyone using \texttt{swh-identify}, an open source tool developed
by Software Heritage, and distributed via PyPI as \texttt{swh.model} (stable
version at 
\swhidref{swh:1:rev:6cab1cc81118877e2105c32b08653509475f3eaa;\\
origin=https://pypi.org/project/swh.model/}).

\subsection{Code fragment}
\label{sec:org9c64b98}
A particularly nifty feature of the SWH-IDs supported by Software Heritage is the
ability to pinpoint a fragment of code inside a specific version of a file, by
using the \texttt{lines=} qualifier available for identifiers that point to
files.

Let's see this feature at work in our running example, which shows clearly
how an article can be greatly enhanced by providing pointers to
code fragments.

\begin{figure}[t]
\centering
    \begin{subfigure}[t]{0.49\textwidth}
        \centering
        \includegraphics[scale=0.5]{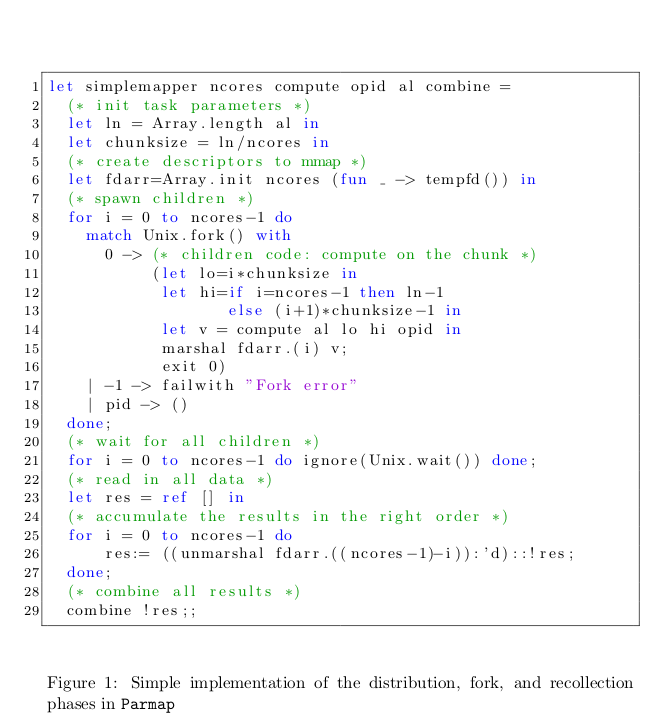}
        \caption{as presented in the article \cite{Parmap2012}}\label{fig:parmappaper}
    \end{subfigure}
    \vspace{1em}
    \begin{subfigure}[t]{0.49\textwidth}
        \centering
        \includegraphics[scale=0.48]{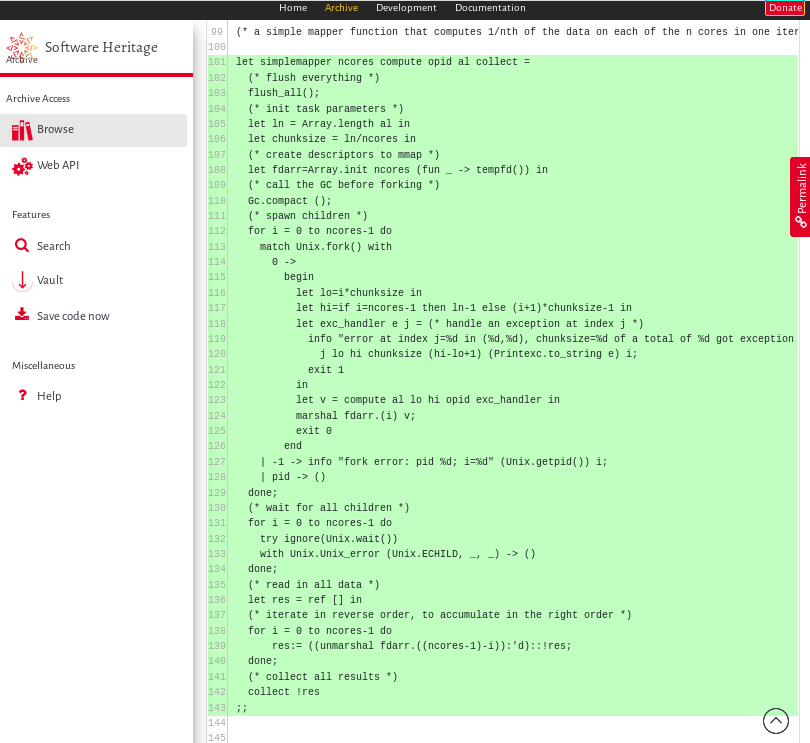}
        \caption{as archived in Software Heritage}\label{fig:parmapswh}
    \end{subfigure}
    \caption{Code fragment from the published article compared to the content
      in the Software Heritage archive}
  \label{fig:parmap}
\end{figure}

In Figure 1 of \cite{Parmap2012}, which is shown here as Figure
\ref{fig:parmappaper}, the authors want to present the core part of the code
implementing the parallel functionality that constitutes the main contribution
of their article. The usual approach is to typeset in the article itself \emph{an
excerpt of the source code}, and let the reader try to find it by delving
into the code repository, which may have evolved in the mean time.
Finding the exact matching code can be quite difficult, as the code excerpt
is \emph{often edited} a bit with respect to the original, sometimes to drop
details that are not relevant for the discussion, and sometimes due to
space limitations.

In our case, the article presented 29 lines of code, slightly edited from the
43 actual lines of code in the Parmap library: looking at
\ref{fig:parmappaper}, one can easily see that some lines have been dropped
(102-103, 118-121), one line has been split (117) and several lines simplified
(127, 132-133, 137-142).\\

Using Software Heritage, the authors can do a much better job, because the original
code fragment can now be precisely identified by the following Software Heritage identifier:
\begin{tcolorbox}
swh:1:cnt:d5214ff9562a1fe78db51944506ba48c20de3379;\\
origin=https://gitorious.org/parmap/parmap.git;\\
lines=101-143
\end{tcolorbox}
This identifier can also be easily obtained using the permalink box shown in Section \ref{ssec:getswhid} above, and it
will \textbf{always} point to the code fragment shown in Figure \ref{fig:parmapswh}.\\

The caption of the original article shown in Figure \ref{fig:parmappaper} can
then be significantly enhanced by incorporating all the clickable links needed
to point to the exact source code fragment that has been edited for inclusion
in the article, as shown in Figure \ref{fig:swhall}.

\begin{figure}[h!]
\begin{tcolorbox}
Simple implementation of the distribution, fork, and recollection phases in \texttt{Parmap} (slightly simplified from the \href{https://archive.softwareheritage.org/swh:1:cnt:d5214ff9562a1fe78db51944506ba48c20de3379;origin=https://gitorious.org/parmap/parmap.git;lines=101-143}{actual code} present in \href{https://archive.softwareheritage.org/swh:1:rev:0064fbd0ad69de205ea6ec6999f3d3895e9442c2;origin=https://gitorious.org/parmap/parmap.git}{the version of Parmap used for this article})
\end{tcolorbox}
\caption{A caption text with links to code fragment and revision}\label{fig:swhall}
\end{figure}

When clicking on the hyperlinked text in the caption shown above, the reader is
brought seamlessly to the Software Heritage archive on a page showing the
corresponding source code archived in Software Heritage, with the relevant
lines highlighted (see Figure \ref{fig:parmapswh}).\\

For \LaTeX{} users, the caption of \ref{fig:swhall} can be written using a few
convenient auxiliary macros, as shown in Figure \ref{fig:swhref}.

\begin{figure}[h!]
\begin{tcolorbox} 
\begin{lstlisting} 
\newcommand{\swhurl}[1]{https://archive.softwareheritage.org/#1}
\newcommand{\swhref}[2]{\href{\swhurl{#1}}{#2}}

...

 \caption{Simple implementation of the distribution,
 fork, and recollection phases in \texttt{Parmap}
 (slightly simplified from the 
 \swhref{swh:1:cnt:d5214ff9562a1fe78db51944506ba48c20de3379;
         origin=https://gitorious.org/parmap/parmap.git;
         lines=101-143}
        {actual code})
 presented in 
 \swhref{swh:1:rev:0064fbd0ad69de205ea6ec6999f3d3895e9442c2;
         origin=https://gitorious.org/parmap/parmap.git}
        {the version of Parmap used in this article}
\end{lstlisting}
\end{tcolorbox}
\caption{Adding clickable hyperlinks to Software Heritage in \LaTeX}\label{fig:swhref}
\end{figure}

\section{Acknowledgements}
\label{sec:orgf6044d3}
These guidelines result from extensive discussions that took place over several
years. Special thanks to Alain Girault, Morane Gruenpeter, Julia Lawall, Arnaud
Legrand and Nicolas Rougier for their precious feedback on earlier versions of this document.

\clearpage



\appendix\clearpage

\section{Appendix: Reference for SWH-ID identifiers}
\label{sec:org36c0b21}
\label{sec:identifiers}

The SWH-ID identifier schema is \href{https://docs.softwareheritage.org/devel/swh-model/persistent-identifiers.html}{fully documented online} and is discussed in the
article \cite{swhipres2018}, but we reproduce here for completeness an excerpt of the
documentation.

\begin{table*}[t]
\caption{EBNF grammar of Software Heritage persistent identifiers}
\label{tab:grammar}
\begin{alltt}
<identifier> ::= "swh" ":" <scheme_version> ":" <obj_type> ":" <obj_id> ;
<scheme_version> ::= "1" ;
<obj_type> ::=
    "snp"  (* snapshot *)
  | "rel"  (* release *)
  | "rev"  (* revision *)
  | "dir"  (* directory *)
  | "cnt"  (* content *)
  ;
<obj_id> ::= 40 * <hex_digit> ;
           (* intrinsic object id, as hex-encoded SHA1 *)
<hex_digit> ::= "0" | "1" | "2" | "3" | "4" | "5" | "6" | "7" | "8" | "9"
              | "a" | "b" | "c" | "d" | "e" | "f" ;
\end{alltt}
\end{table*}

\subsection{Syntax}
\label{sec:org96608b5}
Syntactically, persistent identifiers are generated by the \verb|<identifier>|
entry point of the EBNF grammar given in Table~\ref{tab:grammar}.

\subsection{Semantics}
\label{sec:orgf9e78c6}
The \texttt{swh} prefix makes explicit that these identifiers are related to
Software Heritage, and the colon (\verb|:|) is used as separator between the
logical parts of identifiers. The scheme version (currently \verb|1|) is the
current version of this identifier scheme.

A persistent identifier points to a single object, whose type is explicitly
captured by \verb|<object_type>|:
\begin{description}
\item[snp] identifiers points to snapshots,
\item[rel] to releases,
\item[rev] to revisions,
\item[dir] to directories,
\item[cnt] to contents.
\end{description}

The actual object pointed to is identified by the intrinsic identifier
\verb|<object_id>|, which is a hex-encoded (using lowercase ASCII characters)
SHA1~\cite{SHA1} computed on the content and metadata of the object
itself.\footnote{See
  \url{https://docs.softwareheritage.org/devel/swh-model/persistent-identifiers.html}
  for more details.}

\subsection{Git compatibility}
\label{sec:orgde6bb03}
Intrinsic object identifiers for contents, directories, revisions, and releases
are, at present, compatible with the Git way of computing identifiers for its
objects. A Software Heritage content identifier will be identical to a Git blob
identifier of any file with the same content, a Software Heritage revision
identifier will be identical to the corresponding Git commit identifier,
etc. This is not the case for snapshot identifiers as Git doesn’t have a
corresponding object type. Git compatibility is incidental and is not guaranteed
to be maintained in future versions of this scheme (or Git), but is a convenient
feature for developers, for the time being.

\subsection{Examples}
\label{sec:org571a0e5}
The identifiers below are all interesting examples of what the Software Heritage
identifiers look like.\\

They are resolved by the Software Heritage browsing
pages available at:\\
\texttt{https://archive.softwareheritage.org/<identifier>}

\begin{tcolorbox}
\href{https://archive.softwareheritage.org/swh:1:cnt:94a9ed024d3859793618152ea559a168bbcbb5e2}
{swh:1:cnt:94a9ed024d3859793618152ea559a168bbcbb5e2}
\end{tcolorbox}
points to the content of a file containing the full text of the GPL3 license\\
\begin{tcolorbox}
\href{https://archive.softwareheritage.org/swh:1:dir:d198bc9d7a6bcf6db04f476d29314f157507d505}
{swh:1:dir:d198bc9d7a6bcf6db04f476d29314f157507d505}
\end{tcolorbox}
points to a directory containing the source code of the Darktable photography
application as it was at some point on 4 May 2017\\
\begin{tcolorbox}
\href{https://archive.softwareheritage.org/swh:1:rev:309cf2674ee7a0749978cf8265ab91a60aea0f7d}
{swh:1:rev:309cf2674ee7a0749978cf8265ab91a60aea0f7d}
\end{tcolorbox}points to a commit in the development history of Darktable,
dated 16 January 2017, that added undo/redo supports for masks\\
\begin{tcolorbox}
\href{https://archive.softwareheritage.org/swh:1:rel:22ece559cc7cc2364edc5e5593d63ae8bd229f9f}
{swh:1:rel:22ece559cc7cc2364edc5e5593d63ae8bd229f9f}
\end{tcolorbox}
points to Darktable release 2.3.0, dated 24 December 2016\\
\begin{tcolorbox}
\href{https://archive.softwareheritage.org/swh:1:snp:c7c108084bc0bf3d81436bf980b46e98bd338453}
{swh:1:snp:c7c108084bc0bf3d81436bf980b46e98bd338453}
\end{tcolorbox}
points to a snapshot of the entire Darktable Git repository taken on 4 May 2017
from GitHub.

\begin{table*}[t]
\caption{EBNF grammar of complementary contextual information}
\label{tab:context_grammar}
\begin{alltt}
<identifier_with_context> ::= <identifier> [<lines_ctxt>] [<origin_ctxt>] ;
<lines_ctxt> ::= ";" "lines" "=" <line_number> ["-" <line_number>] ;
<origin_ctxt> ::= ";" "origin" "=" <url> ;
<line_number> ::= <dec_digit> + ;
<url> ::= (* RFC 3986 compliant URLs *)  ;
\end{alltt}
\end{table*}

\subsection{Contextual information}
\label{ssec:context}

It is often useful to complement persistent identifiers with contextual
information about the object's setting.
Currently it is possible to extend the identifier with the optional elements
below using the dedicated syntax presented in Table~\ref{tab:context_grammar}:
\begin{itemize}
\item the software origin where an object has been found/observed
\item the line number(s) of interest, usually within a content object
\end{itemize}

The semi-colon (\verb|;|) is used as a separator between the object identifier
and other contextual information. Each piece of contextual
information is specified as a key/value pair, using the equal sign (\verb|=|)
as a separator.
The extended contextual elements should be added in the following manner:
\begin{description}
\item[software origin] a URL where a given object has been found or observed in
  the wild and used by Software Heritage to ingest the object into the archive.
\item[line numbers] a single line number or a line range, two numbers separated
  with the hyphen (\verb|-|).  Note that line numbers are purely indicative and
  are not meant to be stable, as in some degenerate cases (e.g., text files
  which mix different types of line terminators) it is impossible to resolve
  them unambiguously.
\end{description}

\bigskip

For example, the following identifier \
\begin{tcolorbox}
\href{https://archive.softwareheritage.org/swh:1:dir:c6f07c2173a458d098de45d4c459a8f1916d900f;origin=https://github.com/id-Software/Quake-III-Arena/}
{swh:1:dir:c6f07c2173a458d098de45d4c459a8f1916d900f; \ origin=https://github.com/id-Software/Quake-III-Arena}
\end{tcolorbox}
points to the source code root directory of the computer game Quake III
Arena\footnote{See \url{https://en.wikipedia.org/wiki/Quake_III_Arena}} with
the origin URL where it was found; while \
\begin{tcolorbox}
\href{https://archive.softwareheritage.org/swh:1:cnt:41ddb23118f92d7218099a5e7a990cf58f1d07fa;origin=https://github.com/chrislgarry/Apollo-11;lines=64-72/}
{swh:1:cnt:41ddb23118f92d7218099a5e7a990cf58f1d07fa; \ lines=64-72}
\end{tcolorbox}
points to a comment segment with the warning "NOLI SE TANGERE" in a file in the Apollo-11 source code.\\
\end{document}